\documentclass[useAMS,usenatbib,letterpaper]{mn2e}
\usepackage{times}
\usepackage{amssymb}
\usepackage{lscape,graphicx}

\title[The clustering and abundance of galaxies at
  $z \sim 2$]{The clustering and abundance of star-forming and passive galaxies at $z \sim 2$}
\author[W. G. Hartley et al.]{W.~G. Hartley$^{1}$\thanks{E-mail:
    ppxwh1@nottingham.ac.uk}, K.~P. Lane$^{1}$, O. Almaini$^{1}$, M. Cirasuolo$^{2}$,
  S. Foucaud$^{1}$, C. Simpson$^{3}$, 
  \newauthor S. Maddox$^{1}$, I. Smail$^{4}$, C.~J. Conselice$^{1}$,
  R.~J. McLure$^{2}$, J.~S. Dunlop$^{5}$ \\
  $^{1}$School of Physcis and Astronomy, University of Nottingham, University Park, Nottingham NG7 2RD \\
  $^{2}$SUPA\thanks{Scottish Universities Physics Alliance} Institute for Astronomy, 
  University of Edinburgh, Royal Observatory, Edinburgh EH9 3HJ \\
  $^{3}$Astrophysics Research Institute, Liverpool John Moores University, Twelve Quays House, Egerton Wharf, Birkenhead CH41 1LD   \\
  $^{4}$Institute for Computational Cosmology, Department of Physics, Durham University, Durham DH1 3LE\\
  $^{5}$Department of Physics and Astronomy, University of British Columbia,  6224 Agricultural Rd., Vancouver, B.C., V6T 1Z1, Canada\\
}

\begin{document}

\date{}

\pagerange{\pageref{firstpage}--\pageref{lastpage}} \pubyear{2008}

\maketitle

\label{firstpage}

\begin{abstract}
We use the UKIDSS Ultra-deep survey (UDS), currently the deepest panoramic near infra-red
survey, together with deep Subaru optical
imaging to measure the clustering, number counts and luminosity
function of galaxies at $z\sim 2$ selected using the BzK selection
technique. We find that both star-forming
(sBzK) and passive (pBzK) galaxies, to a magnitude limit of $K_{AB} <
23$, are strongly clustered.  The passive galaxies are the most
strongly clustered population, with scale lengths of $r_0 =
15.0^{+1.9}_{-2.2}$h$^{-1}$Mpc compared with $r_0 =
6.75^{+0.34}_{-0.37}$h$^{-1}$Mpc for star-forming galaxies. The direct
implication is that passive galaxies inhabit the most massive
dark-matter halos, and are thus identified as the progenitors of the
most massive galaxies at the present day.  In addition, the pBzKs
exhibit a sharp flattening and potential turn-over in their number
counts, in agreement with other recent studies. This plateau cannot be
explained by the effects of incompleteness. We conclude that only very
massive galaxies are undergoing passive evolution at this early epoch,
consistent with the downsizing scenario for galaxy evolution. Assuming a purely 
passive evolution for the pBzKs from their median redshift to the present day, their 
luminosity function suggests that only $\sim 2.5 \%$ of present day massive ellipticals 
had a pBzK as a main progenitor.
\end{abstract}


\section{Introduction}

There is growing evidence to support the view that the most massive objects in
the Universe were the first to assemble and complete their star formation
(\citealt{Kodama_etal:2004, Thomas_etal:2005, DeLucia_etal:2004, Bundy_etal:2006, Stott_etal:2007}); this
phenomenon has become known as {\it downsizing} \citep{Cowie_etal:1999}. A number of key issues remain
unresolved however. There are indications that the build up of the galaxy colour 
bimodality occurs around $z\sim2$ (e.g. \citealt{Cirasuolo_etal:2007}), 
but the precise evolutionary path from the distant Universe to the
present day is still undetermined. The mechanism that terminates the major
episode of star-formation is  poorly understood \citep{Benson_etal:2005}, and
it is now clear that massive galaxies undergo significant size evolution from
$z\sim2$ to the present day (e.g. \citealt{Cimatti_etal:2008} and references therein).

Galaxy clustering is an important tool for investigating these populations,
since the amplitude of clustering on large scales ($>1$Mpc) can provide a
measurement of the dark matter halo mass \citep{Mo_and_White:1996, Sheth_and_Tormen:1999}. 
In principle, therefore, one can relate galaxy populations from the
distant past to the present day by tracing the evolution of dark matter halos
within the context of a  framework for structure formation.

A full exploration of these issues will require large spectroscopic surveys of
IR-selected galaxies over a representative volume of the distant
Universe. Such surveys are currently at an early stage, so recent work has
focused on the photometric colour selection of passive vs. star-forming galaxies at
the crucial $z\sim 2$ epoch. The two key methods to date are the BM/BX
selection \citep{Erb_etal:2003}, which is an extension of the Lyman-break dropout
technique, and the BzK technique \citep{Daddi_etal:2004} which is based on
$K$-band selection.  Recent studies have shown that the BM/BX technique is
reasonably efficient at selecting actively star-forming galaxies at $z\sim 2$,
but largely misses the passive galaxy population at these epochs
\citep{Quadri_etal:2007, Grazian_etal:2007}. In contrast, the BzK method appears
to be the most complete of the broad-band techniques for the selection of both
star-forming and passive galaxies \citep{Daddi_etal:2004, VanDokkum_etal:2006, 
Grazian_etal:2007}. 
Based on initial $K$-band selection, in principle it
does not suffer the same heavy biases caused by dust or the age of the stellar
populations.

An alternative is to use a larger set of filters over a range of wavelengths and infer 
a galaxy's redshift from comparison of the calculated magnitudes with a set of 
templates (\citealt{Cirasuolo_etal:2007} and references therein). This work makes use of 
such photometric redshifts and in principle 
one can derive the stellar age of a galaxy using such templates. However 
due to the uncertainties in determining stellar ages from the template fits, and 
the desire for comparison with the literature, the bulk of the present analysis is 
based on simple BzK selection.

The BzK technique was first developed using the K20 survey
\citep{Cimatti_etal:2002} by \cite{Daddi_etal:2004}, and preferentially selects 
star-forming and passive
galaxies in the redshift range $1.4 < z < 2.5$ by using the B-, z- and
K$_s$-band broadband filters. Recent work has shown that the number counts of
these populations differ markedly \citep{Kong_etal:2006, Lane_etal:2007}, with
star-forming galaxies being far more abundant at all
magnitudes. The abundance of sBzK galaxies allowed \cite{Kong_etal:2006}
(henceforth K06) to perform a detailed clustering analysis, segregated by
limiting K-band luminosity in the range $18.5 < K_{vega} < 20$. They found
that the clustering of sBzKs is strongly dependent on $K$-band luminosity.
\cite{Hayashi_etal:2007} studied the clustering of sBzKs over a smaller area
(180 arcmin$^{2}$) but to much greater depth (K$_{AB} < 23.2$) and confirmed
this strong luminosity dependence. The sBzK population therefore appear to 
inhabit a range of halo masses, and are therefore likely to be the progenitors
of a wide range of present day galaxies.

There have been fewer studies of {\em passive} galaxy clustering at high
redshift, in part because of their relatively low surface density.  K06 found
early evidence that pBzKs are more strongly clustered than sBzKs, although the
limited depth and area of this survey ($320 {\rm arcmin}^2$ to K$_{Vega}=20$) gives
rise to significant statistical uncertainly, particularly given the relatively
low surface density of pBzKs at these magnitudes ($0.38$ arcmin$^{-2}$). \cite{Blanc_etal:2008} 
measured the clustering and number counts of BzK-selected galaxies to the same depth 
as K06, but over a much 
wider area (0.71 deg$^2$ total between two fields). Their BzK clustering amplitudes are smaller than those of K06 
but consistent at the $1\sigma$ level. Analysis of
deeper survey data over a similar area is required to confirm these results, and we 
present such an analysis 
in this work.

\cite{Lane_etal:2007} (henceforth L07) used the Early Data Release (EDR) from
the UKIDSS UDS to investigate the number counts and overlap in colour-selected
galaxies, including those selected by the BzK technique. They found that the flattening in pBzK
number counts observed by K06 extends to become a plateau at the EDR depth of
$K_{AB} < 22.5$ and noted that if this were to continue to turn over it could
imply an absence of high-redshift passive galaxies at low luminosities.

In this work we investigate the number counts and clustering of BzK-selected galaxies 
over a substantially greater depth and area than any previous
study. Section 2 describes the data and galaxy number counts. In Section 3 we
present an angular clustering analysis, which is extended by de-projection to
infer the real-space correlation lengths. In Section 4 we use photometric
redshifts to derive a luminosity function for these populations, and compare
with the present-day galaxy luminosity function. A
discussion and conclusions are presented in Section 5. Throughout this paper we 
assume a flat $\Lambda$CDM cosmology with $\Omega_m = 0.3$ and $h = 0.73$.

\section[]{Source identification and number counts}

\subsection{Data set and sample definitions}

\begin{figure}
\begin{center}
\includegraphics[angle=0, width=240pt]{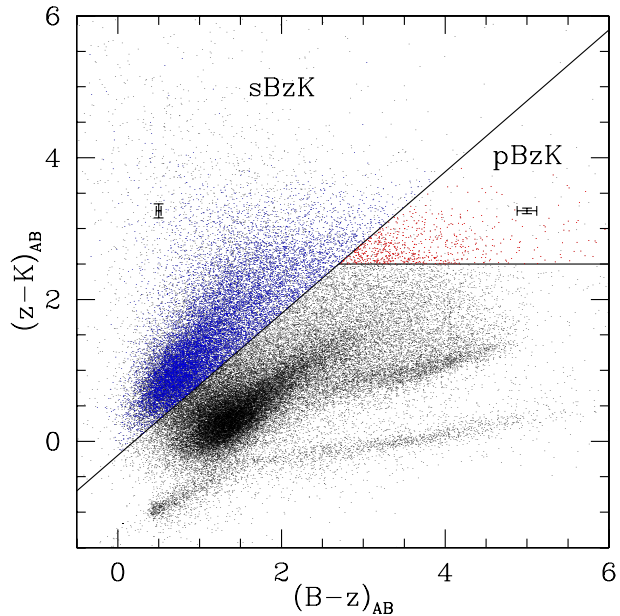}
\caption{BzK colour-colour diagram for sources in the UDS DR1. The sBzK and pBzK selection regions are marked accordingly. Also visible are the passive galaxy track identified in \protect\cite{Lane_etal:2007}, and the stellar locus used to match our photometric system to that of \protect\cite{Daddi_etal:2004} (see text). We show mean error-bars for the galaxies in the BzK selection regions.}
\label{bzk}
\end{center}
\end{figure}

The UKIRT Infrared Deep Sky Survey (UKIDSS) has been underway since spring 2005
and comprises 5 sub-surveys covering a range of areas and depths
\citep{Lawrence_etal:2007}. This survey has been made possible by the advent of the
UKIRT Wide-Field Camera \citep{Casali_etal:2007}.

We base the present study on data from the deepest component of
UKIDSS, the Ultra Deep Survey (Almaini et al. in preparation). This
aims to reach final depths of $K_{AB} = 25.0, H_{AB} = 25.4, J_{AB} =
26.0$ ($5\sigma$, point source, 2'') over an area of 0.8 deg$^2$. The size
of the UDS field significantly reduces the effects of cosmic variance
and on this scale is the deepest near-infrared survey to date. For
this work we use the UDS DR1 release \citep{Warren_etal:2007}, which
reaches $5\sigma$ (point source) depths of K$_{AB} = 23.5$ and J$_{AB} = 23.7$. For
details of the completeness estimation, image stacking, mosaicing and
catalogue extraction procedures see \cite{Foucaud_etal:2007} and Almaini et
al.  (in preparation). In addition to these data, deep
$B,V,R,i^{\prime},z^{\prime}$ imaging is also available from Subaru
Suprimecam to limiting depths of $ B_{AB} = 28.4, ~V_{AB} = 27.8, ~R_{AB} = 27.7,
~i_{AB}^{\prime} = 27.7$ and $ z_{AB}^{\prime} = 26.7$ \citep{Furusawa_etal:2008}. 
The UDS and Subaru survey areas are not entirely
coincident, which reduces our usable area to 0.63 deg$^2$ in this
analysis.

\subsection{Construction of the BzK colour-colour diagram}

The original BzK selection technique was introduced in \citet{Daddi_etal:2004}
based on photometry in the Bessel $B$-band, Gunn $z$-band and $K_s$-bands as
defined on the FORS1, FORS2 and ISAAC instruments at the Very Large Telescope
(VLT).  To correct our colours to these filters we used the stellar track of
K06, who in turn used published empirical stellar spectra from \citet{Pickles:1998} and
synthetic stellar spectra from \citet{Lejeune_etal:1997}, convolved with the
filter responses of the VLT instruments used in \citet{Daddi_etal:2004}. These derived colours
provide a convenient reference for matching stars on the well-defined BzK
stellar locus. We use the adjusted stellar locus of K06 
in the following way.

The stellar locus can be split into a main branch and a ``knee''
feature, clearly visible in the lower part of figure \ref{bzk}. The intersection of 
these branches provides a fixed point 
in the BzK plane from which we can derive
a quantitative adjustment. Using a least squares fit to each section, we derived the following
offset in $z-K$ and $B-z$ colours to this fixed point to
convert to the photometric system of \citet{Daddi_etal:2004}:
\begin{equation} 
(z-K)_{Daddi} = (z^{\prime}-K)_{UDS} - 0.26
\end{equation} 
\begin{equation} 
(B-z)_{Daddi} = (B-z^{\prime})_{UDS} + 0.06 
\end{equation}

This correction is then applied to all sources in the
BzK plane. We henceforth refer to BzK photometry defined in this way.

The standard BzK definitions of \citet{Daddi_etal:2004} were then used to
construct our BzK sample.  K-band sources ($> 5 \sigma$) from the UDS were
used as the primary catalogue, with optical magnitudes extracted directly from
the Subaru imaging data after careful matching of the Subaru and UDS
astrometric frames. Aperture magnitudes were then extracted using a 2-arcsec
diameter. Only sources outside the contaminated halos of saturated
optical stars were used.

\begin{figure}
\begin{center}
\includegraphics[angle=0, width=240pt]{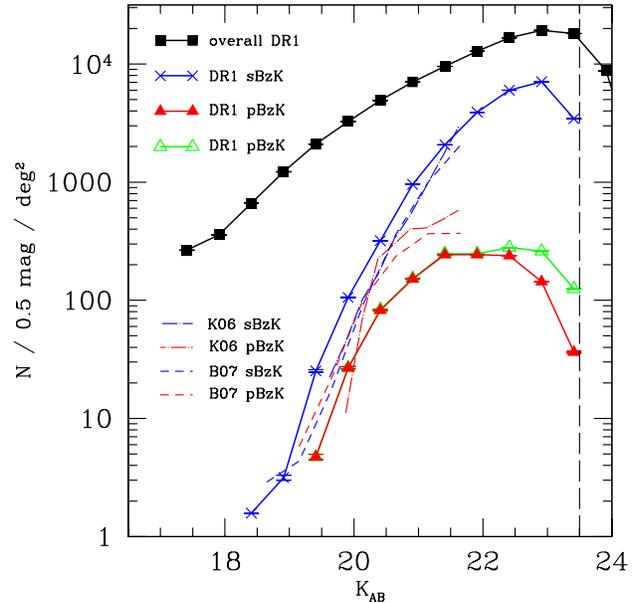}
\caption{Differential number counts of passive and star-forming BzK galaxies with apparent
K-band magnitude, compared with the full-sample of $K$-selected
galaxies. The error-bars shown are standard Poisson errors on the counts. Two sets of 
values are shown for pBzKs: the standard selection of pBzKs (filled, red triangles); and 
the worst case sample described in the text (open, green triangles). Numerical values 
are reproduced in table \protect\ref{tab-counts}.  Literature values from K06 and 
\protect\cite{Blanc_etal:2008} are also shown 
(dot-dashed lines and dashed lines respectively). The plateau identified in previous 
works (K06, L07, \protect\citealt{Blanc_etal:2008}) 
and subsequent turn-over prior to our magnitude limit are apparent in the pBzK number counts.}
\label{counts}
\end{center}
\end{figure}

All K-band sources with $K_{AB} < 23.5$ were used for
BzK selection unless both the B- and z$^{\prime}$-band magnitudes were fainter than the $3
\sigma$ limits measured on those images ($0.29\%$ of sources outside of 
contaminated regions), since these cannot be constrained
within the BzK plane.

Overall this procedure results in the selection of 15177 sBzKs (21.7\% of the full sample) 
and 742 pBzKs (1.06 \% of the sample),
11551 and 702 respectively at $K_{AB} < 23.0$.

\subsection{Number counts}

\begin{figure}
\begin{center}
\includegraphics[angle=0, width=240pt]{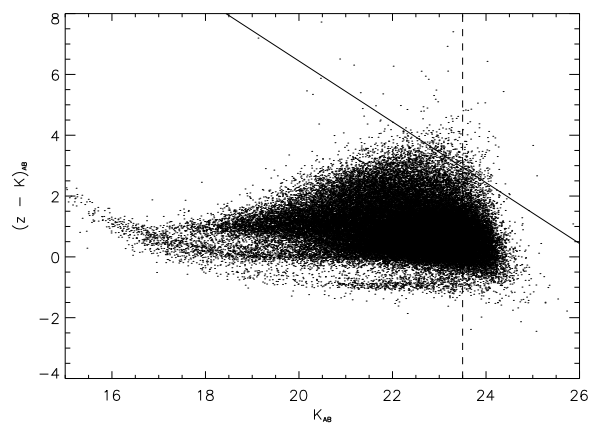}
\caption{$(z - K)_{AB}$ colour versus K-band magnitude. The dashed line is our completeness 
limit of $K_{AB} = 23.5$ for the full sample, while the solid line shows how 
the completeness limit 
in z affects $(z - K)$ colour. We begin to be incomplete above $(z - K) > 2.94$. The impact of 
this incompleteness is discussed in the text.}
\label{kzk}
\end{center}
\end{figure}

\begin{table}
\caption{Differential Number Counts in log(N/deg$^{2}$/0.5mag) bins
  for sBzK and pBzK in the UDS DR1.}
\begin{tabular} {|l|l|l|l|l|}
\hline
K bin&all&sBzK&pBzK&Worst\\
centre&sources& & &case pBzK\\
\hline
17.41&2.832&0.198&-&-\\
17.91&2.803&0.676&-&-\\
18.41&2.957&0.198&-&-\\
18.91&3.203&0.500&-&-\\
19.41&3.380&1.403&0.676&0.676\\
19.91&3.556&2.025&1.429&1.429\\
20.41&3.720&2.504&1.914&1.923\\
20.91&3.872&2.982&2.181&2.185\\
21.41&3.998&3.315&2.386&2.394\\
21.91&4.121&3.592&2.386&2.394\\
22.41&4.229&3.778&2.377&2.431\\
22.91&4.285&3.849&2.158&2.394\\
23.41&4.258&3.536&1.560&2.158\\
23.91&3.942&0.801&-&-\\
\hline
\end{tabular}
\label{tab-counts}
\end{table}

The differential K-band number counts are illustrated in Figure \ref{counts}, and
tabulated in Table 1, for both the full sample of galaxies and those selected
as passive and star-forming BzKs. We find a steep rise in the counts of
star-forming sBzKs toward fainter magnitudes.  In contrast, pBzK number counts
exhibit an apparent flattening at $K\sim 21$, consistent with the findings of
K06 and L07. Since these galaxies are sampled over a relatively narrow
redshift range, this may imply that the passive population consists largely of
luminous galaxies at this early epoch (see Section 4).

To investigate the reality of the turn-over, we note that the galaxies around
and beyond this feature have very faint B-band magnitudes, with a substantial
fraction (62\%) below the $3\sigma$ detection limit in this band ($B_{lim} = 28.4$).  
A non-detection
in the $B$-band does not affect their classification, however, and merely
pushes them red-ward in B-z colour in the BzK diagram illustrated in Figure
1. The more worrying class are the objects which are below the $3\sigma$ limit 
in $z'$ {\em
and} $B$, which cannot be assigned BzK classification (164 objects). Figure \ref{kzk} 
shows that even within 
our K-band limit, we become incomplete above $(z - K) \simeq 3$. We study the effect 
of this incompleteness 
by considering the extreme case in which all of the objects with B- 
and $z^{\prime}$-band magnitudes 
fainter than the $3\sigma$ limits (28.4 and 26.7 respectively) are pBzKs. In this 'worst case' 
sample the number counts still exhibit a plateau, as before, but with the absence of a turn-over.

At our conservative limit of $K_{AB}< 23$ our completeness is $> 95$\% for compact 
galaxies (Almaini et al., in prep.), so the feature is unlikely to be due to  $K$-band
incompleteness unless there is unprecedented size-evolution in the pBzK
population towards fainter magnitudes compared to the general population.  The
extreme compactness of passive galaxies at these redshifts observed by \cite{Cimatti_etal:2008} 
suggests that of the two classes, the sBzKs should suffer more from such incompleteness. 
There is no evidence from their number counts to suggest that they are 
adversely affected in this 
way, so we expect that the pBzKs are likewise unaffected below this magnitude.

Photometric errors are another source of concern.  Adjusting the
 boundaries for pBzK selection by the mean photometric errors for 
sources close to the boundaries ($0.1$ magnitudes),
 we found that such errors had no
 noticeable impact on the decline in the pBzK population. Additionally, since the
 pBzK region of the colour-colour plane is sparsely populated,
 photometric errors are likely to scatter fainter objects {\em into}
 pBzK selection, rather than the opposite.  The faint pBzK counts are
 thus likely to be slightly {\em overestimated} because of photometric
 errors.

We conclude that the flattening in the pBzK
population is likely to be real and the subsequent turn-over is probable but not certain. 
We note that the turnover corresponds to
absolute magnitude $ M_K = -23.6$ at z = 1.4, which is close to the value of
$L^*$ determined for the K-band luminosity function at these redshifts
\citep{Cirasuolo_etal:2007}.  Taken at face value, these results are
suggestive of a sharp decline in the number density of pBzK galaxies towards
lower luminosities, consistent with expectation from downsizing.  As correctly 
pointed out in \cite{Grazian_etal:2007} and
predicted by \cite{Daddi_etal:2004}, this is not necessarily equivalent to a
true decline in the number density of passive galaxies, since the efficiency
and completeness of the pBzK technique is largely untested at such faint
magnitudes. In particular, there is some evidence that passive galaxies can
also be found among the redder sBzK galaxies (at large values of $z-K$),
with a possible incompleteness as high as $34\%$ \cite{Grazian_etal:2007}. 
We find that randomly adding an additional $34$\% of galaxies to the pBzK sample
from this part of the diagram is indeed sufficient to remove the apparent
turn-over, since these are predominantly the fainter sBzK galaxies.

The sBzK number counts are significantly higher than those of K06, \cite{Blanc_etal:2008} 
and \cite{Imai_etal:2008} (not plotted); while the pBzK number counts are lower. The UDS field is 
more than 6 times larger than that of the K20 survey and 4 times larger than the one used in 
\cite{Imai_etal:2008}. The combined field used in \cite{Blanc_etal:2008} is of a similar size to the UDS,
cosmic variance is the most likely explanation for the difference in number counts in this case.

\section[]{Clustering properties}

\subsection{Angular clustering}

The 2-point angular correlation function, $w(\theta)$ is defined by the joint 
probability of finding two galaxies in solid angle elements $\delta\Omega_1$ 
and $\delta\Omega_2$ at a given separation \citep{Peebles:1980},
\begin{equation}
\delta P = n^2 \delta\Omega_1 \delta\Omega_2 (1 + w(\theta_{12})).
\end{equation}

\begin{figure}
\begin{center}
\includegraphics[angle=0, width=240pt]{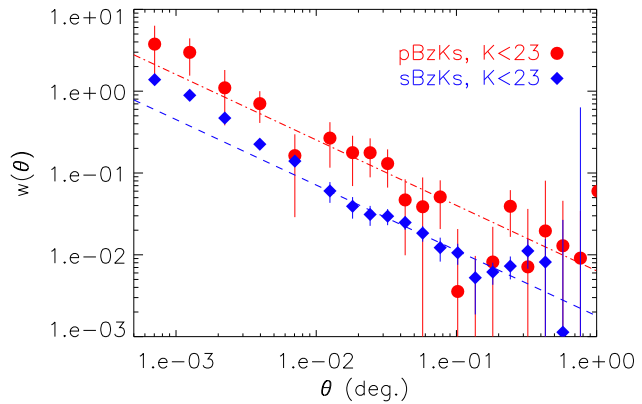}
\caption{The angular correlation function for our BzK-selected galaxy
samples. The best-fitting power laws are shown, with slopes fixed to the
fiducial value of $\delta=0.8$. The pBzK galaxies are very strongly clustered, much more 
so than sBzKs, indicating that they occupy the most massive dark matter halos at 
their epoch.}
\label{wth}
\end{center}
\end{figure}

To estimate the correlation function we use the estimator of \cite{Landy_and_Szalay:1993},

\begin{equation}
w(\theta) = \frac{DD - 2DR + RR}{RR}
\end{equation}

where DD, DR and RR are the counts of data-data, data-random and
random-random pairs respectively at angular separation $\theta$,
normalised by the total number possible. Although this estimator is
relatively robust against systematic errors, there remains a small bias due
to the finite field size, which is corrected for an integral
constraint by a constant, C. We follow the method of \cite{Roche_and_Eales:1999} by using
the random--random counts to estimate the size of this bias:

\begin{equation}
C=\frac{\Sigma N_{RR}(\theta)\theta^{-0.8}}{\Sigma N_{RR}(\theta)},
\end{equation}

where the sums extend to the largest separations within the field.

The sBzK and pBzK samples were selected to a limit of $K_{AB} <
23$. We adopted this conservative magnitude limit to ensure the
minimum contamination by spurious sources.
We fit a single power law of the form $w(\theta) = A(\theta^{-\delta} - C)$ to
the corrected data over the separation range $0.01 - 0.1$ degrees, fixing the
slope to the fiducial $\delta=0.8$ and minimising the $\chi^2$. The errors on the measurements 
are a combination of those due to shot noise (estimated by bootstrap resampling) and an estimate 
of those due to cosmic variance. The estimates for cosmic variance were found by splitting the field 
into 4 and computing the variance of $w(\theta)$.

In order to fit the clustering reliably, we wish 
to avoid small-scale excesses due to multiple galaxy occupation of a single halo.
The lower bound was chosen to correspond to $\sim 0.9$Mpc (co-moving) at z$\sim 2$ for 
this reason. The upper bound is a conservative estimate of the limit to which our data have 
enough signal for a reliable fit to be obtained. 

For sBzK galaxies the amplitude was found
to be $A=1.79^{+0.17}_{-0.17} \times 10^{-3}$ (deg.$^{0.8}$)and for the pBzK
population $A = 6.37^{+1.58}_{-1.54} \times 10^{-3}$ (deg.$^{0.8}$). The clustering 
amplitude of the pBzKs is inconsistent 
with the clustering of sBzK galaxies at the 3-sigma level. Figure 
\ref{wth} shows the clustering measurements corrected for the integral constraint 
for the sBzK and pBzK galaxies.

\subsection{De-projected clustering amplitude}

The real space clustering and projected clustering are linked by the
relativistic Limber equation \citep{Limber:1954}.  If the redshift
distribution of a sample is known, the Limber equation can be inverted and the
correlation length, $r_0$, can be calculated in a robust manner \citep{Peebles:1980,
Magliocchetti_and_Maddox:1999}

\begin{figure}
\begin{center}
\includegraphics[angle=0, width=240pt]{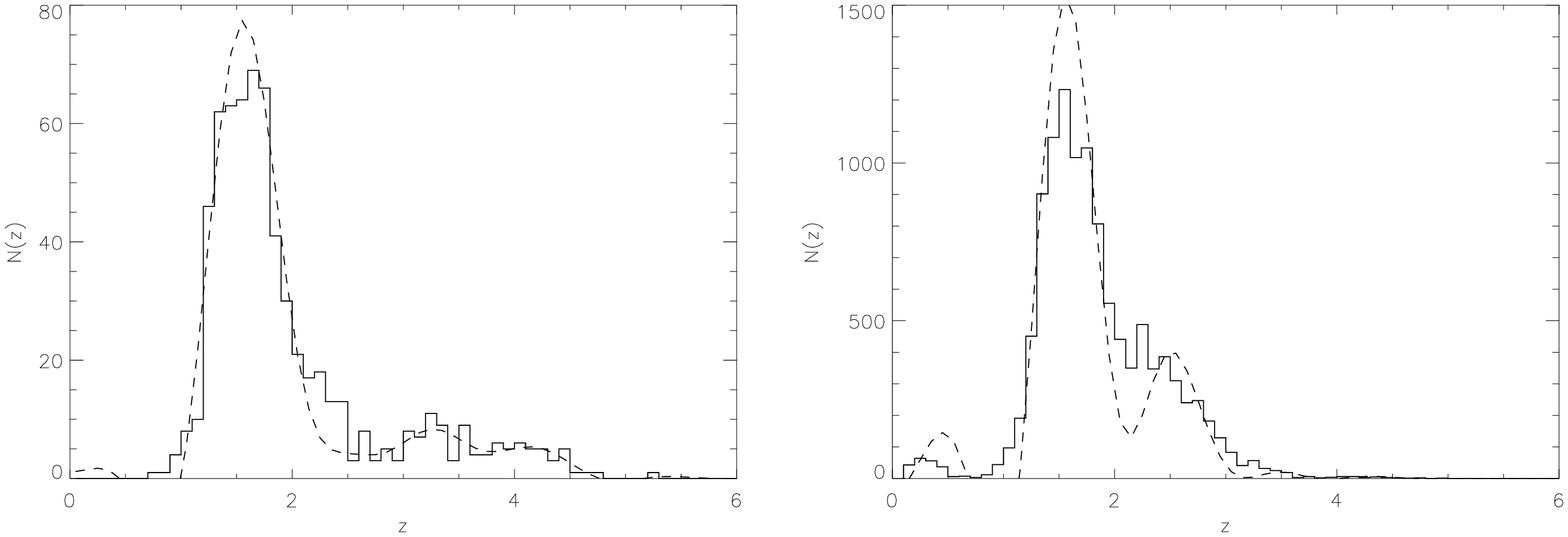}
\caption{Photometric redshift distributions for passive (left) and star-forming (right) BzK-selected 
galaxies (solid line histogram). The 
over-plotted dashed lines are distributions with photometric errors de-convolved (see text).}
\label{nz}
\end{center}
\end{figure}

To estimate the redshift distribution we use photometric redshifts based
on the Subaru and UDS bands previously introduced, with the addition of
Spitzer data taken as part of the SWIRE survey
\citep{Lonsdale_etal:2003}. The method is described fully in
\cite{Cirasuolo_etal:2007}, and consists of minimising the $\chi^2$ of
synthetic galaxy templates. The distributions are shown in figure \ref{nz}.
The photometric redshift distribution was then used directly in the inverted Limber's equation 
during the calculation of $r_0$. The values obtained in this manner are 
r$_0 = 15.8^{+2.0}_{-2.2}h^{-1}$ Mpc and $7.69^{+0.40}_{-0.41}h^{-1}$ Mpc for $K < 23$
pBzKs and sBzKs respectively. The quoted errors are due to the error in the fit to the  
clustering amplitude and therefore take into account the shot noise and cosmic variance.

The photometric redshifts are subject to errors ($\sigma/(1+z) = 0.095$ and $0.105$ for pBzKs and sBzKs 
respectively), however, and assuming that the true redshift distribution is 
highly peaked, these errors are likely to broaden the measured distribution. A broader distribution will 
result in a larger inferred clustering scale length. It is therefore important that we take such errors 
into account. We do so by assuming the errors are Gaussian, and then deconvolve the errors from the redshift 
distribution using a Fourier-based Wiener filter. A more detailed description of this method is provided in 
the appendix. The resulting de-convolved redshift distributions are shown in figure \ref{nz}.

The scaling lengths recovered using the corrected redshift distribution are as follows: 
$15.0^{+1.9}_{-2.2}h^{-1}$ Mpc and $6.75^{+0.34}_{-0.37}h^{-1}$ Mpc for pBzKs and sBzKs respectively.

However, we note that our
photometric redshift code is largely untested for pBzK and sBzK
galaxies. This work 
will be refined by the use of an ongoing ESO Large
Programme using VIMOS and FORS2 on the VLT to target one sixth of the 
UDS DR1 galaxies with photometric redshifts $> 1$. 

\begin{figure}
\begin{center}
\includegraphics[angle=0, width=240pt]{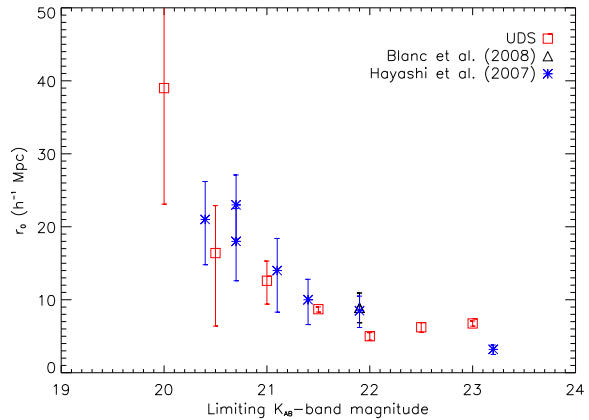}
\caption{The dependence of clustering strength of sBzK-selected galaxies on 
limiting K$_{AB}$-band magnitude. Our measurements (open squares) are shown
together with literature values from \protect\cite{Blanc_etal:2008} and 
\protect\cite{Hayashi_etal:2007} (open triangle and asterisks respectively). Our values 
confirm the magnitude dependence of sBzK clustering strength.}
\label{sBzK}
\end{center}
\end{figure}

We also confirm previous claims for a
dependence of sBzK clustering on apparent magnitude
\citep{Hayashi_etal:2007}. Figure \ref{sBzK} and Table \ref{sBzKtab} show the values for 
sBzKs with varying limiting magnitude. The dependence is much stronger 
at magnitudes of $K < 22$, indicating a strong correlation between halo mass and sBzK 
luminosity for these objects.

\begin{table}
\caption{Values of r$_0$ for sBzKs in the UDS by limiting K$_{AB}$-band magnitude 
along with the number of objects brighter than the relevant limit. As for the full samples, the 
error ranges are derived from estimates of the shot noise and cosmic variance.}
\begin{tabular} {|l|l|l|}
\hline
K$_{AB,lim}$&N&r$_0$\\
\hline
20.0&92&$39.0^{+11.8}_{-15.9}$\\
20.5&250&$16.4^{+6.5}_{-10.0}$\\
21.0&689&$12.6^{+2.7}_{-3.2}$\\
21.5&1724&$8.71^{+1.49}_{-1.68}$\\
22.0&3789&$5.00^{+0.49}_{-0.55}$\\
22.5&7199&$6.23^{+0.64}_{-0.69}$\\
23.0&11551&$6.75^{+0.34}_{-0.37}$\\
\hline
\end{tabular}
\label{sBzKtab}
\end{table}

\subsection{BzK galaxies selected at$1.4 < z < 2.5$ }

The BzK selection was defined by \cite{Daddi_etal:2004} to isolate galaxies in the redshift 
range $1.4 < z < 2.5$. Clearly from figure \ref{nz}, and as expected by \cite{Daddi_etal:2004}, 
there are contaminating objects from outside of this range.  Using our photometric redshifts 
we can asses how successful the BzK selection technique is in reproducing the clustering of 
objects within the desired range.

Following the same method outlined for the full samples, we compute clustering amplitudes 
find correlation lengths of 
$11.1^{+1.7}_{-1.8}h^{-1}$ Mpc and $5.46^{+0.36}_{-0.37}h^{-1}$ Mpc for $1.4 < z < 2.5$ pBzKs and sBzKs 
respectively. These values are slightly lower than those for the full samples and indicate that the 
high redshift tails in the full sample are at least as highly clustered as those within the $1.4 < z < 2.5$ range.
However, the conclusions drawn from the full sample are still valid, namely that the pBzK galaxies are 
significantly more strongly clustered than the sBzK galaxies.

\section{Luminosity function}

\begin{figure*}
\begin{minipage}{150mm}
\begin{center} 
\includegraphics[angle=0, width=\textwidth]{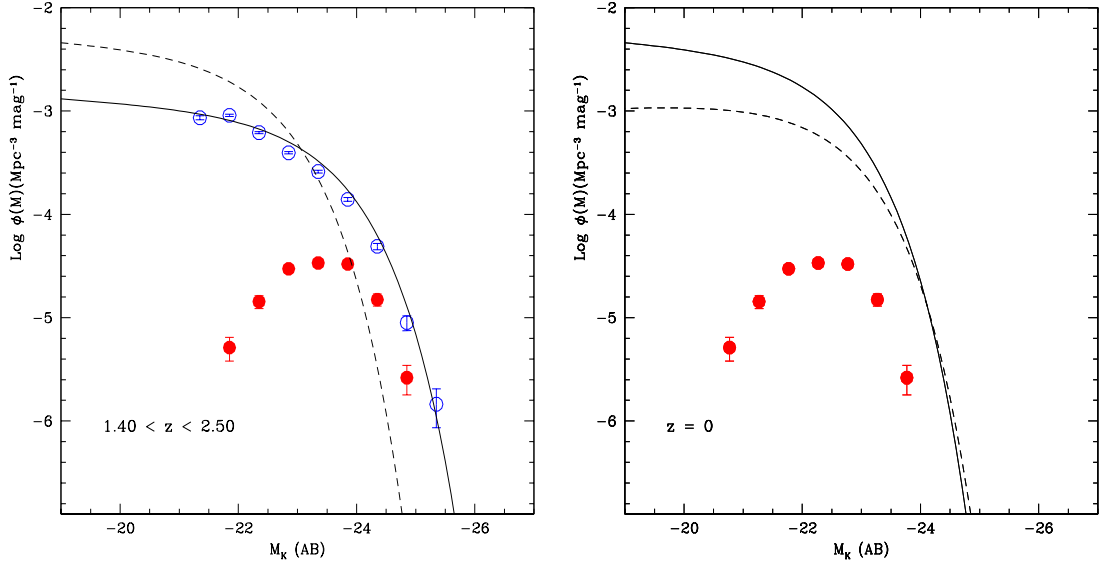}
\caption{{\em Left}: Luminosity functions for our pBzK and sBzK samples (filled and open 
circles respectively), plus all
$K$-selected galaxies to a limit of $K < 23$ (restricted to galaxies
with photometric redshifts in the range $1.4 < z <2.5$, solid line). The sBzKs are
representative of the overall population in the range $1.4 < z < 2.5$,
while  the pBzKs are found to be exclusively bright objects. Also shown is the $z=0$ 
luminosity function for K-selected galaxies from \protect\cite{Kochanek_etal:2001} (dashed line).
{\em Right}: 
The solid line shows the same luminosity function for $z=0$ galaxies as the dashed line in 
the right hand panel. The dashed curve
shows the $z=0$ luminosity function for early type galaxies from \protect\cite{Kochanek_etal:2001}, 
while the points are the $z=0$ luminosity function for pBzKs under the assumption of 
passive evolution (see text).}
\label{lum}
\end{center}
\end{minipage}
\end{figure*}

A luminosity function was constructed for our BzK-selected galaxies in the same way 
as detailed in \cite{Cirasuolo_etal:2007}, by using the $1/V_{max}$ method 
\citep{Schmidt:1968}.
Figure \ref{lum} shows the luminosity function for the BzK galaxies with photometric redshifts 
in the range $1.4<z<2.5$ (points with error-bars),
compared with all K-selected galaxies in the same range (solid line). Also plotted is 
the $z=0$ luminosity function from \cite{Kochanek_etal:2001}. It is clear that sBzK
galaxies sample a wide range in luminosity, while the pBzK population
are dominated by bright galaxies with $M_K>-23$. At this epoch,
however, the bright end of the luminosity function is still dominated
by star-forming objects, consistent with \cite{Cirasuolo_etal:2007}
who found that the galaxy colour bimodality is weak at these
redshifts. 
The pBzK galaxies are nevertheless likely to be among the
most massive systems, since the mass-to-light ratio will be
significantly lower for the actively star-forming systems.

Under the assumption that the pBzKs passively evolve to the present
day (with minimal merging) one can estimate their contribution to the
bright end of the present-day luminosity function. The implied
evolution is modelled from $z\sim 1.60$ (the median value for pBzKs) 
to $z=0$ assuming the spectral evolution models of
\cite{Bruzual_and_Charlot:2003}.  For simplicity we chose a solar
metalicity model and a Salpeter initial mass function \citep{Salpeter:1955} and assume only
passive evolution from initial formation bursts at redshifts $z_f=3$
and $z_f=10$.  This results in a Johnson $K$-band absolute magnitude
evolution in the range $0.96 < \Delta M_{K} < 1.21$. Taking the mean
value we can estimate the minimum number of present day
bright ($K < -23.4$) galaxies that could previously have been pBzKs. Under our assumptions 
we find that $2.5\%$ of such galaxies can be explained by passively evolved pBzKs. It is 
assumed that the remainder is made up of the descendants of bright sBzKs and merger remnants from 
within and between the two classes.

\section[]{Discussion and conclusions}

We present the number counts, clustering and luminosity function of galaxies at $z\sim 2$
selected using the BzK selection criteria. The pBzK galaxies show a
marked flattening in their number counts which cannot be explained by the effects
of incompleteness, and a possible turn over at
faint magnitudes. We conclude that, at this epoch, it is generally 
luminous, massive galaxies that are undergoing passive evolution. This is consistent
with the down-sizing scenario, in which the most massive galaxies are
formed first and are the first to evolve onto the red sequence \citep{Kodama_etal:2004}.

The angular clustering of the passive galaxy sample is very strong,
approximately $4$ times the amplitude of the sBzK population.
This is in part due to their relatively narrow redshift distribution
(K06), but a large difference remains after de-projection to the
real-space correlation length. We find $r_0 = 15.0^{+1.9}_{-2.2}h^{-1}$Mpc and
$r_0 = 6.75^{+0.34}_{-0.37}h^{-1}$Mpc for the pBzK and sBzK galaxies
respectively. Our value for the correlation length of sBzKs is almost twice that 
found by \cite{Hayashi_etal:2007}, who used a sample with similar limiting magnitude 
but with smaller field size ($180$ arcmin$^2$, compared with $\sim 2250$ arcmin$^2$ for the UDS). Our 
sample consists of more than 10 times the number 
of sBzKs than that of \cite{Hayashi_etal:2007} and in addition we make fewer assumptions regarding 
their redshift distribution. Further surveys reaching depths of 
$K_{AB,lim} \sim 23$ are required to fully account the effects of cosmic variance, however. 

We also confirm that the scale-length for the clustering
of sBzK galaxies is dependent on apparent magnitude, consistent with
the work of \cite{Hayashi_etal:2007} (figure \ref{sBzK}). This dependence is 
far more significant at magnitudes 
below $K_{AB} \sim 22$ indicating a strong correlation between such galaxies and 
the mass of the hosting halo.

In addition to this luminosity dependence there is a clear enhancement at small 
scales in the sBzK clustering. This enhancement is indicative of multiple sBzK 
galaxies occupying a single dark matter halo. The scale at which this turn-off 
occurs can provide an indication of the size, and hence mass, 
of the hosting dark matter halos. The turn-off occurs at $\sim 0.01$ deg, which 
corresponds to $\sim 0.3$ Mpc at z$=1.65$. Halos of this size have masses 
between $10^{11}$ and $10^{12} M_{\sun}$ \citep{Mo_and_White:2002}. A full consideration 
of this enhancement within the halo occupation distribution framework will be 
presented in a future paper. The host halo mass 
can also be derived from comparing the de-projected clustering scale length with 
models of dark matter halo clustering evolution. Using models based on 
\cite{Mo_and_White:2002} we find a typical halo mass of $\sim 6\times10^{11} M_{\sun}$.

Applying the model of clustering evolution to the pBzKs
we can qualitatively conclude that they are inhabitants of the most massive
halos at their epoch (in excess of $10^{13} M_{\odot}$); halos which
will eventually become massive groups and clusters by the present day. The evolutionary path that 
pBzKs take with redshift will also be the subject of future work, using further colour 
selection techniques.

Our conclusion based on the $r_0$ measurement is strengthened by their luminosity function. Even 
under the strict assumption of purely passive evolution, the descendants of pBzKs 
occupy the bright end of the luminosity function. Such galaxies are group and cluster 
dominant galaxies in the local universe. The brightest ($K < 21$) sBzKs have clustering 
scale lengths comparable to, or greater than, that which we have found for the pBzKs. This 
finding indicates that such galaxies will also become group and cluster dominant 
galaxies by $z=0$. The implication is that we are 
indeed witnessing the epoch at which the build up of the red sequence begins.

\section*{Acknowledgments}

OA, IRS and RJM acknowledge the support of the Royal Society.  SF, MC, KPL and WGH acknowledge 
the support of STFC. We are
grateful to Xu Kong for help in matching photometric
filters and also to Kaz Sekiguchi and Hisanori Furusawa for the Subaru data used in this study.  
We also extend out gratitude to the staff at UKIRT for their
tireless efforts to ensure that the UKIDSS surveys are a success. We would also like to thanks the anonymous referee for their thorough reading and useful comments.

\section*{Appendix}

In this Appendix we describe the Fourier method used to deconvolve photometric redshift errors from 
the n(z) distribution, which can then be used to invert the Limber equation. The errors in the 
photometric redshift distribution are assumed to be Gaussian and it is also assumed that 
the measured distribution is simply the convolution of these errors with the true redshift distribution. Under 
these assumptions it should be possible to deconvolve the errors from the measured distribution, using the 
fact that a convolution is simply a multiplication in the Fourier domain. Dividing the Fourier transform of 
the measured distribution by that of the Gaussian errors, should then give us an estimate of the 
true distribution.

In practice, a relatively large number of terms in the discreet Fourier transform are required to reproduce the 
redshift distribution accurately, and as the the Fourier transform of a Gaussian is also a Gaussian, small 
levels of high frequency noise can be amplified greatly to give a spurious result. One way to avoid such a 
problem is to use a Wiener filter:

\begin{equation}
W(k) = \frac{1}{H(k)}\times\left(\frac{H(k)^2}{H(k)^2+\frac{1}{SNR}}\right).
\end{equation}

Where H(k) is the Fourier transform of the Gaussian errors and SNR the signal to noise ratio. In the limit of 
SNR being infinite, this filter tends to $1/H(k)$ as in the simple deconvolution above. SNR is 
estimated by fitting the power spectrum of the redshift distribution to a function of the form 
$a\times10^{-b.k} + c$, with c identified as being the noise level. The resultant redshift distributions are 
shown in figure \ref{nz}.

\bibliographystyle{mn2e.bst}
\bibliography{mn-jour,papers_cited_by_KL}

\label{lastpage}

\end{document}